\documentclass[sigconf]{acmart}

\AtBeginDocument{%
  \providecommand\BibTeX{{%
    \normalfont B\kern-0.5em{\scshape i\kern-0.25em b}\kern-0.8em\TeX}}}

\copyrightyear{2023}
\acmYear{2023}
\setcopyright{acmlicensed}\acmConference[CHI '23]{Proceedings of the 2023 CHI Conference on Human Factors in Computing Systems}{April 23--28, 2023}{Hamburg, Germany}
\acmBooktitle{Proceedings of the 2023 CHI Conference on Human Factors in Computing Systems (CHI '23), April 23--28, 2023, Hamburg, Germany}
\acmPrice{15.00}
\acmDOI{10.1145/3544548.3581416}
\acmISBN{978-1-4503-9421-5/23/04}

\newcommand{\add}[1]{\textcolor{black}{#1}}

\begin{document}

\title{Measuring Categorical Perception in Color-Coded Scatterplots}

\author{Chin Tseng}
\affiliation{%
   \institution{University of North Carolina at Chapel Hill}
   \city{Chapel Hill}
   \state{NC}
   \country{USA}}
\email{chint@cs.unc.edu}

\author{Ghulam Jilani Quadri}
\affiliation{%
   \institution{University of North Carolina at Chapel Hill}
   \city{Chapel Hill}
   \state{NC}
   \country{USA}}
\email{jiquad@cs.unc.edu}

\author{Zeyu Wang}
\affiliation{%
   \institution{University of North Carolina at Chapel Hill}
   \city{Chapel Hill}
   \state{NC}
   \country{USA}}
\email{zeyuwang@cs.unc.edu}

\author{Danielle Albers Szafir}
\affiliation{%
   \institution{University of North Carolina at Chapel Hill}
   \city{Chapel Hill}
   \state{NC}
   \country{USA}}
\email{danielle.szafir@cs.unc.edu}

\begin{abstract}
Scatterplots commonly use color to encode categorical data. However, as datasets increase in size and complexity, the efficacy of these channels may vary. Designers lack insight into how robust different design choices are to variations in category numbers. This paper presents a crowdsourced experiment measuring how the number of categories and choice of color encodings 
used in multiclass scatterplots 
influences the viewers’ abilities to analyze data across classes. 
Participants estimated relative means in a series of scatterplots with 2 to 10 categories encoded using ten color palettes 
drawn from popular design tools. 
Our results show that the number of categories and color 
discriminability within a color palette 
notably impact people's perception of categorical 
data in scatterplots 
and that the judgments become harder as the number of categories grows. We examine existing palette design heuristics in light of our results to help designers make robust color choices informed by the parameters of their data.
\end{abstract}



\maketitle

\section{Introduction}
\label{sec-intro}
%
%
Scatterplots 
enable people to conduct a wide variety of 
statistical tasks~\cite{quadri2021survey}, such as finding outliers \cite{saket2018task, li2010size}, estimating mean values~\cite{gleicher2013perception, hong2021weighted, kim2018assessing}, and 
assessing correlation~\cite{yang2018correlation,rensink2010perception}. 
Multiclass scatterplots leverage people's abilities to attend to different subsets of information in order to compare patterns across different categories of data. 
When the number of categories 
increases, 
people's abilities to analyze patterns across categories may degrade~\cite{blasius1998visualization}.
However, certain scatterplot designs may be more robust to larger numbers of categories than others. 
Determining robustness is challenging as the perception of multiclass scatterplots 
requires 
first identifying points from relevant categories and then estimating values from those points across various visual encodings~\cite{gleicher2013perception}.

Existing studies provide guidance for supporting a range of tasks in
general scatterplots~\cite{quadri2021survey}, such as cluster estimation~\cite{sedlmair2012taxonomy,quadri2020modeling,quadri2022automatic}, 
or for tuning across channels such as 
color differences~\cite{szafir2018modeling} and point size~\cite{hong2021weighted,li2010size}.
However, little attention has been paid to 
design choices for rendering complex multiclass scatterplots and how such design choices may change as the number of categories increases.
Color palettes~\cite{munzner2014visualization, stone2014engineering} and shapes~\cite{burlinson2017open} are commonly used to delineate categories in scatterplots, but available design guidance for effectively supporting categorical tasks is largely heuristic rather than empirical, which raises questions as to the robustness and precision of this guidance across a range of scenarios.

Gleicher et al.~\cite{gleicher2013perception} and Burlinson et al.~\cite{burlinson2017open}
offer preliminary experimental insight into the robustness of different visual channels on 
mean estimation in multiclass scatterplots. However, these studies focus on scatterplots with two to three classes, where as we 
measure the effect of 2--10
categories across different
color palettes. 
The number of categories heavily impacts people's abilities to reason across categories~\cite{haroz2012capacity}, especially for color, which remains the default channel for encoding categorical data in many popular commercial applications~\cite{shmueli2011data,mackinlay2007show}. 
Existing studies~\cite{healey1996choosing, gramazio2016colorgorical, wang2018optimizing} 
highlight both the importance and complexity of selecting proper color palettes for categorical visualizations. Despite the popularity of using color palettes in categorical visualizations, we lack insight into how robust these palettes are to the number of presented categories and as to whether that robustness varies across different parameters of a palette. Such information is critical for effectively communicating data, especially as the size and complexity of data continue to grow. 

We conducted a crowdsourced study 
to measure how robust different palette designs are to increasing numbers of categories. 
Participants estimated mean values in a series of multiclass scatterplots with varying numbers of categories (2--10), dataset sizes (10--20 points per category), and color palettes drawn from popular visualization tools.
We found that both the number of categories and choices of color palettes significantly impact people's abilities to estimate category means. 
We deconstructed our results with respect to common parameters of color encodings to find potential cues for robust palette design and find preliminary evidence that subitizing may impact categorical estimates~\cite{phillips1977components, haroz2012capacity,nothelfer2017redundant,nothelfer2019measures}.

\noindent \textbf{Contribution:} The primary contribution of our paper is evaluating 
mean estimation in multiclass scatterplots with varying color palettes.
Our results 
characterize the effect of the numbers of categories and color palette in perceptions of multiclass scatterplots.
Our findings challenge current guidelines on multiclass scatterplot design~\cite{gleicher2013perception}, and we present an exploratory analysis of key factors for effective color palette design.  

\section{Related Work}
\label{sec-related}

Visual encodings in multiclass scatterplots significantly affect people's ability to interpret categorical data correctly.
However, we still do not understand the perceptual impact of encoding choices across varying numbers of categories.
We briefly review the topics of graphical perception in scatterplots, color palette design, and tasks in scatterplots to ground our work.

\subsection{Graphical Perception in Scatterplots}

Understanding categorical perception is a fundamental task in both cognitive science~\cite{harnad2003categorical} and visualization~\cite{munzner2014visualization}. Past work has introduced a range of techniques for eliciting patterns in categorical data, such as Flexible Linked Axes~\cite{kosara2006parallel}, Parallel Sets~\cite{lex2010comparative}, and Matchmaker~\cite{claessen2011flexible}.
However, these techniques leverage specialized approaches with high learning costs, 
making them difficult for lay audiences to work with.
Scatterplots, alternatively, are more familiar for many audiences and commonly encode categorical data~\cite{sarikaya2018scatterplots}. Consequently, 
understanding how to best design scatterplots for categorical datasets is essential for effective data communication.

Graphical perception studies investigate how effectively people can estimate different properties from visualized data (see Quadri \& Rosen~\cite{quadri2021survey} for a survey). 
Scatterplots are commonly used in graphical perception experiments as they are sufficiently complex to reflect real-world challenges and simultaneously sufficiently simple to control~\cite{rensink2014prospects, rensink2010perception, harrison2014ranking, kay2016beyond}.   
Existing studies have analyzed how scatterplots can support a variety of perceptual tasks across a range of channels. 
For example, Kim \& Heer use scatterplots as a means to assess how different visual channels support 
various tasks~\cite{kim2018assessing}.
Hong et al.~\cite{hong2021weighted} found that varying point size and lightness can lead to perceptual bias in mean judgments in scatterplots. Scatterplot studies commonly investigate how design influences people's abilities to estimate aggregate statistics, such as correlation~\cite{harrison2014ranking,rensink2010perception,kay2016beyond}, clustering~\cite{quadri2022automatic,sedlmair2012taxonomy,quadri2020modeling}, and means~\cite{hong2021weighted,gleicher2013perception,wei2019evaluating,whitlock2020graphical}. 
Other studies model the influence of different channels on scatterplot design, such as opacity~\cite{micallef2017towards}, color~\cite{szafir2018modeling}, and shape~\cite{burlinson2017open}. 

Most graphical perception studies focus on statistical relationships within a single category of scatterplots. However, studies of multiclass scatterplots often characterize people's abilities to separate classes by measuring just-noticeable differences in categorical encodings~\cite{smart2019measuring,burlinson2017open}. Alternatively, Gleicher et al.~\cite{gleicher2013perception} studied how different categorical encodings influenced people's abilities to compare the means of different classes 
with varying numbers of points and
differences in means, colors, and shapes.
They found that scatterplots can effectively reveal interclass differences and that the design of a scatterplot influenced people's abilities to compare classes, with color being the strongest categorical cue. However, in contrast to other work on categorical visualization~\cite{haroz2012capacity}, they found that increasing the number of classes from two to three did not decrease performance.  
We build on these observations to explore how robust people's estimates are in scatterplots with between 2 and 10 classes with varying hardness levels, color palettes, and numbers of points, (see Section \ref{sec-methodology}) to more deeply understand factors involved in effective multiclass scatterplot design. 

\subsection{Color Palette Design}
Gleicher et al.'s findings about the effectiveness of color in multiclass scatterplots echo existing design guidance and results from other studies of categorical data encodings~\cite{gleicher2013perception,haroz2012capacity,trumbo1981theory,munzner2014visualization}.
Choosing a proper categorical color palette\footnote{We define a color \emph{palette} as a set of colors specifically designed for categorical data.} for visualizing categorical data is a crucial task~\cite{trumbo1981theory, zhou2015survey}. Designers employ a combination of color models and heuristics to generate palettes (see Zhou \& Hansen~\cite{zhou2015survey}, Kovesi~\cite{kovesi2015good}, Bujack et al.~\cite{bujack2017good}, and Nardini et al.~\cite{nardini2019making} for surveys).
A range of studies has explicitly examined color perception for continuous data, such as characterizing limitations of rainbow colormaps \cite{ware1988color,reda2020rainbows,borland2007rainbow,quinan2019examining}, comparing the task-based effectiveness of continuous colormap designs~\cite{padilla2016evaluating,reda2018graphical,liu2018somewhere}, modeling color discrimination~\cite{ware2018measuring}, examining color semantics~\cite{anderson2021affective}, quantifying the impact of size and shape on encoding perception~\cite{smart2019measuring, szafir2018modeling} and examining perceptual biases~\cite{schloss2018mapping}.
However, significantly fewer studies have characterized color use for categorical data encoding.  


Several principles and metrics of effective color palette design have been proposed~\cite{brewer1994guidelines,harrower2003colorbrewer,stone2006choosing,gramazio2016colorgorical}. 
Past work recommends that color palettes optimize the mapping between data semantics and color semantics~\cite{lin2013selecting,schloss2020semantic,setlur2016linguistic}; select colors that emphasize color harmonies~\cite{stone2006choosing,zeileis2009escaping}, affect~\cite{bartram2017affective}, or pair preference~\cite{schloss2011aesthetic}; and maximize perceptual and categorical separability between colors~\cite{healey1996choosing} (see Silva et al. \cite{silva2011using} for a survey).
Designers can use predefined metrics to describe aesthetic 
(e.g., pair preference~\cite{schloss2011aesthetic}), perceptual (e.g., 
CIEDE 2000~\cite{sharma2005ciede2000}), 
and categorical (e.g., color name difference or uniqueness~\cite{heer2012color})
attributes of color to implement these guidelines and constrain effective palette design. 
While these metrics underlie many palette design guidelines, implementing these guidelines effectively takes significant expertise. 


Several methods for creating effective color palettes have been introduced.
For example, Healey~\cite{healey1996choosing} considers linear separability, color difference, and color categorization to design discriminable color palettes.
Harrower and Brewer~\cite{harrower2003colorbrewer} introduced ColorBrewer for providing designer-crafted distinguishable color palettes for cartography.
Gramazio et al.~\cite{gramazio2016colorgorical} developed Colorgorical, which can generate categorical palettes by optimizing several perceptual and aesthetic metrics. 
Recent efforts have also explored how palettes might be extracted from images~\cite{zheng2022image} or colors from a given palette optimally assigned to a visualization~\cite{lee2012perceptually, lin2013selecting,wang2018optimizing}. Tools such as Colorgorical~\cite{gramazio2016colorgorical} and ColorBrewer~\cite{harrower2003colorbrewer} enable people to generate or choose from a range of palette designs (see Zhou \& Hansen~\cite{zhou2015survey} for a survey).  
In this study, we compare preconstructed palettes from a range of sources, 
including ColorBrewer~\cite{harrower2003colorbrewer}, 
Tableau~\cite{tableau}, D3~\cite{6064996}, Stata Graphics~\cite{statagraphics19}, and Carto~\cite{carto} (see \autoref{fig:palettes} for the details of our selected color palettes). Following the model for comparing the effectiveness of continuous color ramps in Liu \& Heer \cite{liu2018somewhere}, we leverage these palettes to understand how effectively common best-practice color palettes encode data over a range of data parameters.

\section{Methodology}
\label{sec-methodology}

We analyzed how the number of categories, number of points, and color palettes used to distinguish various categories impact people's abilities to reason with multiclass scatterplots.
We performed a crowdsourced study measuring how well people were able to compare category means over varying category numbers and color palette designs. This study allowed us to characterize the effect of category number in multiclass scatterplots as well as how robust different color palette designs are across varying numbers of categories. 
We hypothesized that:

\begin{itemize}
\item[\textbf{H1:}] \textbf{Performance will decrease as the number of categories increases.}

As visual information becomes more complex, perception and cognition degrades~\cite{liberman1957discrimination, regier2009language}. Haroz \& Whitney~\cite{haroz2012capacity} found that these findings generalized to categorical visualizations: increasing the number of categories degrades visual search performance. However, Gleicher et al.'s findings contradicted this observation, instead finding no performance difference between two or three category visualizations~\cite{gleicher2013perception}.  
We expect that for larger numbers of categories, this robustness will likely falter, even with designer-crafted palettes. \add{Existing heuristics recommend that visualizations should not use more than seven colors for reliable data interpretation \cite{munzner2014visualization}. This guidance suggests that we should see drastic performance reductions for seven or more categories.}

\item[\textbf{H2:}] \textbf{The choices of the color palette will affect people's abilities to effectively compare means.}

Perceptual studies demonstrate that color is a strong cue in both visualization~\cite{szafir2018modeling} and categorical perception~\cite{goldstone1995effects}. Past work has shown that, even in unitless data, the choice of color palettes can affect visualization interpretation~\cite{gramazio2016colorgorical,healey1996choosing}. We likewise anticipate that color palette design may differently support varying numbers of categories: some palettes may more robustly distinguish a range of classes than others, especially as the complexity of the palette increases with larger numbers of colors. 



\end{itemize}

The anonymized data, results, and infrastructure 
for our study can be found on \href{https://osf.io/wz8eb/?view_only=03db060f94ee42f29f453ed3013e3405}{OSF.}\footnote{https://osf.io/wz8eb/?view\_only=03db060f94ee42f29f453ed3013e3405}

\subsection{Task}


Scatterplots have been studied across a range of tasks (see Sarikaya \& Gleicher~\cite{sarikaya2018scatterplots} for a survey). We employed a relative mean judgment task as applied in previous studies~\cite{gleicher2013perception,hong2021weighted,kramer2017visual}. As in Gleicher et al.~\cite{gleicher2013perception}, we asked participants to estimate the category with the highest average y-value. We used this task as it required participants to first find data points of different categories and then estimate statistical values 
over all points in that category. This task is sensitive to both overinclusion (i.e., including points that are not in a given class) and underinclusion (i.e., failing to include points in a given category), meaning that confusion between points of different categories should be reflected in participants' responses. It also represents a basic statistical quantity that most lay participants 
are able to compute.

\begin{figure*}[htbp] 
\vspace{-1em}
\centering
\includegraphics[width=0.7\textwidth]{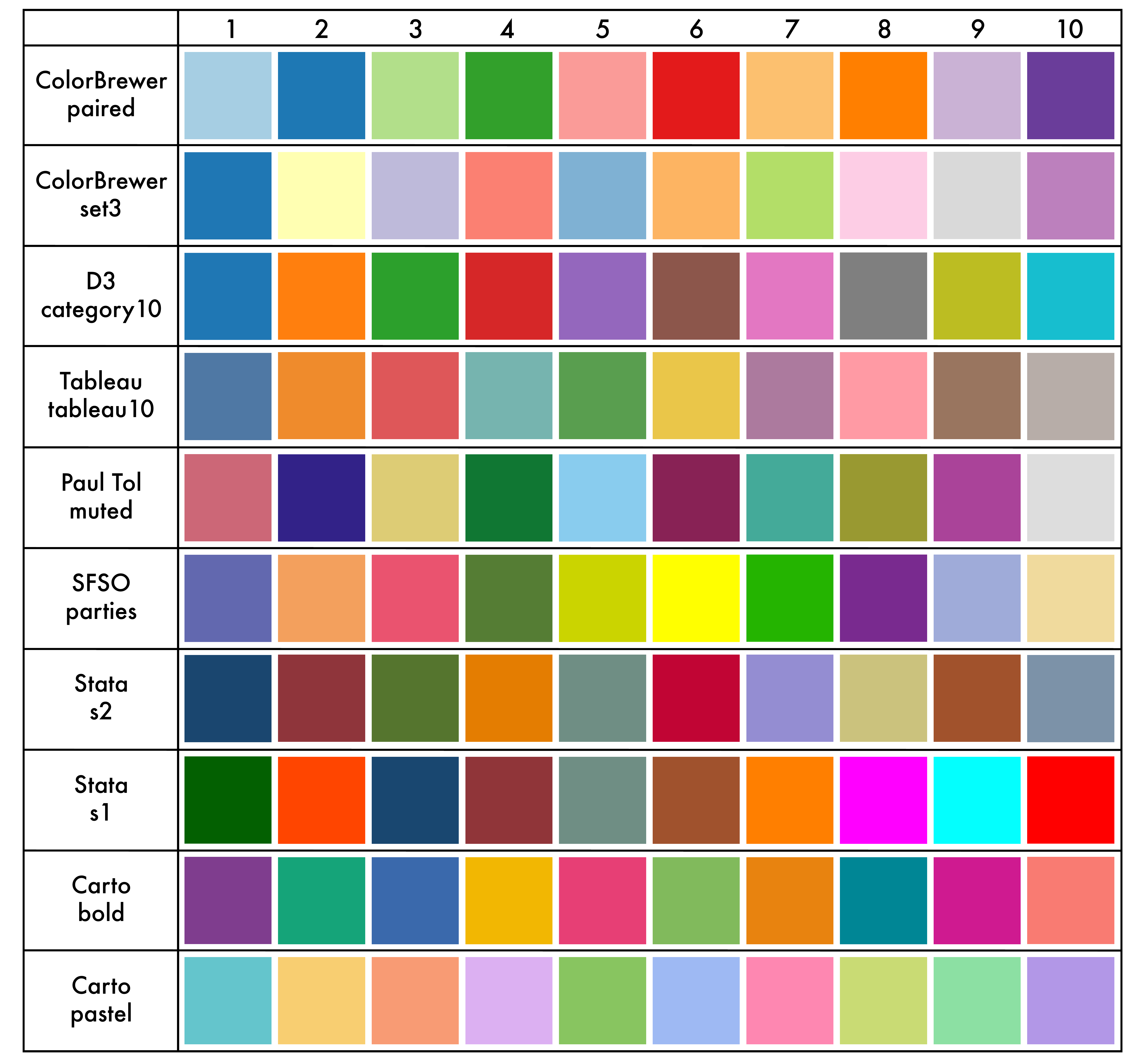} 
\vspace{-1em}
\caption{The 10 color palettes used in our experiment.} 
\label{fig:palettes}
\end{figure*}

\begin{figure*}[htbp] 
\centering
\includegraphics[width=0.8\textwidth]{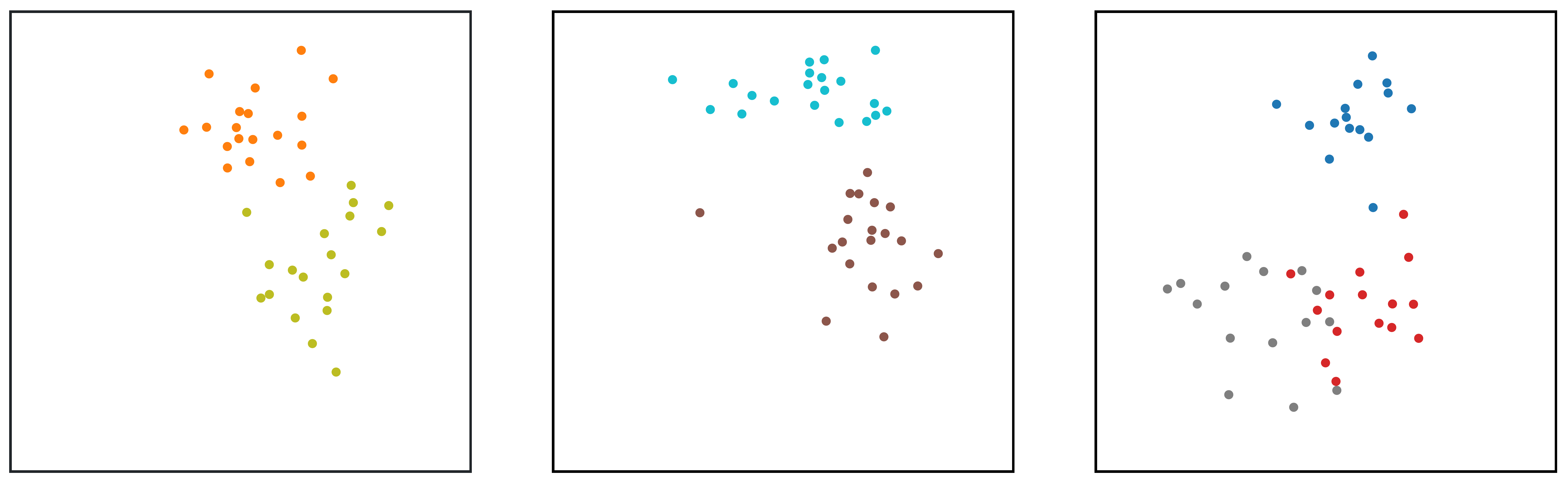} 
\caption{Three engagement checks with D3 color palettes. Participants were required to pass two out of these three tasks to be considered as an approved response. All engagement checks were placed in random order with other formal trials.} 
\label{fig:amt-engagement}
\end{figure*}



\subsection{Stimuli Generation}
\label{sec-stimuli-generation}

\begin{table*}[htbp] 
\centering
\caption{The experiment parameters. We refined the factors and domain range in three pilot studies. Category number and color palettes are our independent variables, and hardness level and point number are the control variables. The experiments were built from the combination of these four factors.
} 
\includegraphics[width=0.9\textwidth]{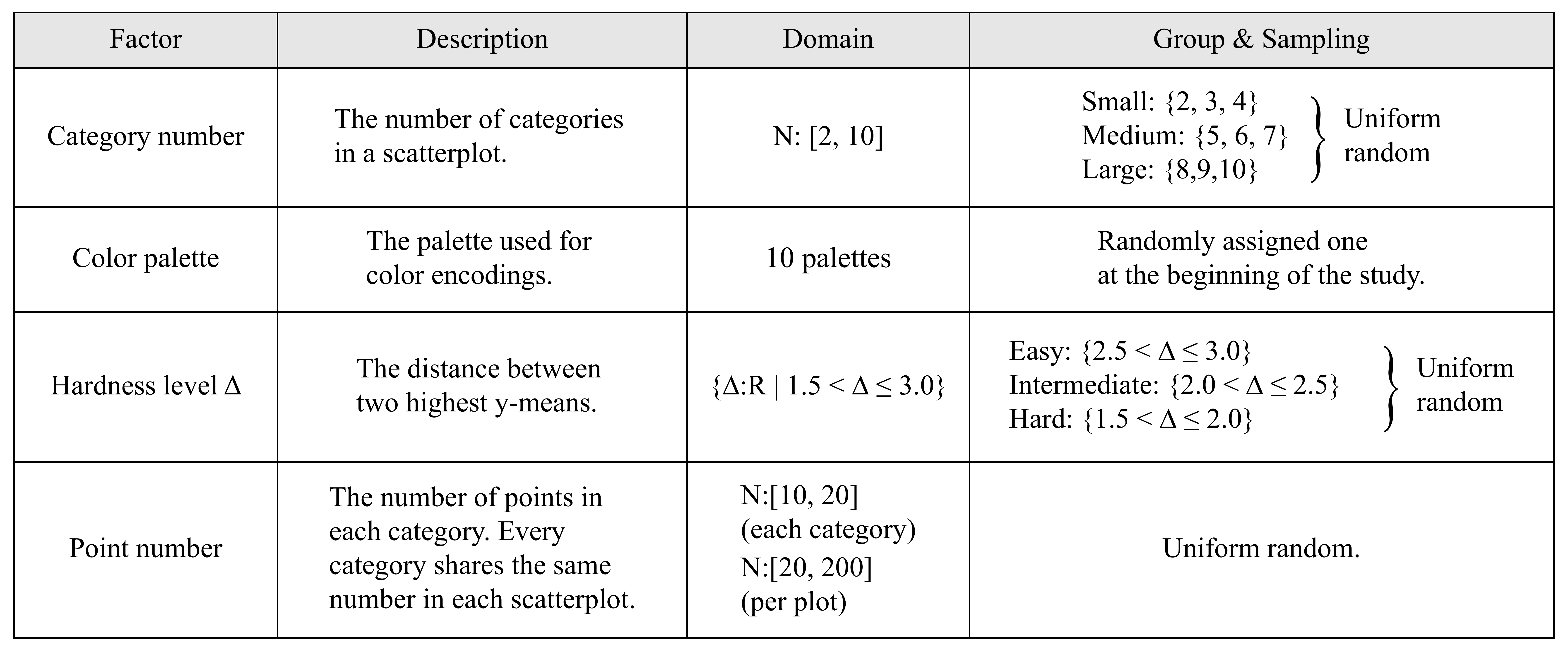} 
\label{tab:parameters}
\end{table*}

Participants estimated means for a series of scatterplots. We generated each scatterplot as a 400x400 pixel graph using D3~\cite{6064996}. Each scatterplot was rendered to white background and two orthogonal black axes with 13 unlabeled ticks. For every point, we 
rendered a filled circle mark with a three pixel radius.
\add{We selected three pixel points based on internal piloting to ensure that points were distinguishable between classes while also minimizing the need to address overdraw and reflecting design parameters commonly seen in real-world visualizations. }



As shown in \autoref{fig:palettes}, we selected 10 qualitative color palettes: ColorBrewer/Paired~\cite{harrower2003colorbrewer}, ColorBrewer/Set3~\cite{harrower2003colorbrewer}, D3/Category10~\cite{6064996}, Tableau/Tab10~\cite{tableau}, Paul Tol/Muted~\cite{tol2012colour}, SFSO/Parties~\cite{sfso}, \newline  Stata/S1~\cite{statagraphics19}, Stata/S2~\cite{statagraphics19}, Carto/Bold~\cite{carto} and Carto/Patel~\cite{carto}. These color palettes were chosen from popular visualization tools that provide 
at least 10 categorical colors in a single palette. If there were more than 10 colors in a certain palette, we used the first 
10 as the palette's colors.
In each scatterplot, colors were randomly selected from the target palette and 
mapped to corresponding categories. While some tools prescribe a fixed order to the selection of colors from a palette, this is not a universal design practice. 
\add{Randomization helps avoids potential bias from differences beyond color selection as not all palettes may have been intentionally ordered, but future work should investigate differences in the ordered application of palettes.}

We tuned our dataset parameters in a series of three extensive pilot studies, measuring performance for varying numbers of categories, points, and hardness levels (see Appendix for details). As in Gleicher et al.~\cite{gleicher2013perception}, we controlled task hardness using the distance between classes. The hardness level is denoted by $\Delta$ and is calculated by the 
distance between y-means of classes in multiclass scatterplots. 
To generate 
positional data with the given mean and covariance, we used a function from Numpy~\cite{oliphant2006guide} that randomly samples from a multivariate normal distribution. We denoted our data points as \begin{math} \{ x,y \in \mathbb{R} \, | \, 0<x,y<10 \} \end{math}. First, we randomly sample the mean \begin{math} \mu{_1} \end{math} in the range [5, 9] for the category that possesses the highest mean, then set the mean \begin{math} \mu{_2} = \mu{_1} - \Delta\end{math} as the second highest mean based on y-values. To prevent subsequent means from drifting too far apart and artificially simplifying the task, we constrained the mean \begin{math} \mu_i \end{math} of the rest of the categories to \begin{math} \Delta < \mu{_1} - \mu_i < 1.5\Delta \end{math}. Finally, we determined the covariance for each category that has y-mean \begin{math} \mu_i \end{math} with \begin{math} cov(\lambda_i,  \lambda_i) \end{math} where \begin{math}\lambda_i=random(1, min(\mu_i, 10-\mu_i))\end{math}. We used this variance to tune the datasets such that selecting the category with the highest point did not reliably produce the correct answer, with variance tuned in piloting.  

Each scatterplot contained between 10 and 20 points per category. 
To prevent overlapping points, we applied jittering methods which add random noise to any data points that would otherwise overlap each other.
We generated 450 datasets in total.

\begin{figure*}[htbp] 
\centering
\includegraphics[width=0.8\textwidth]{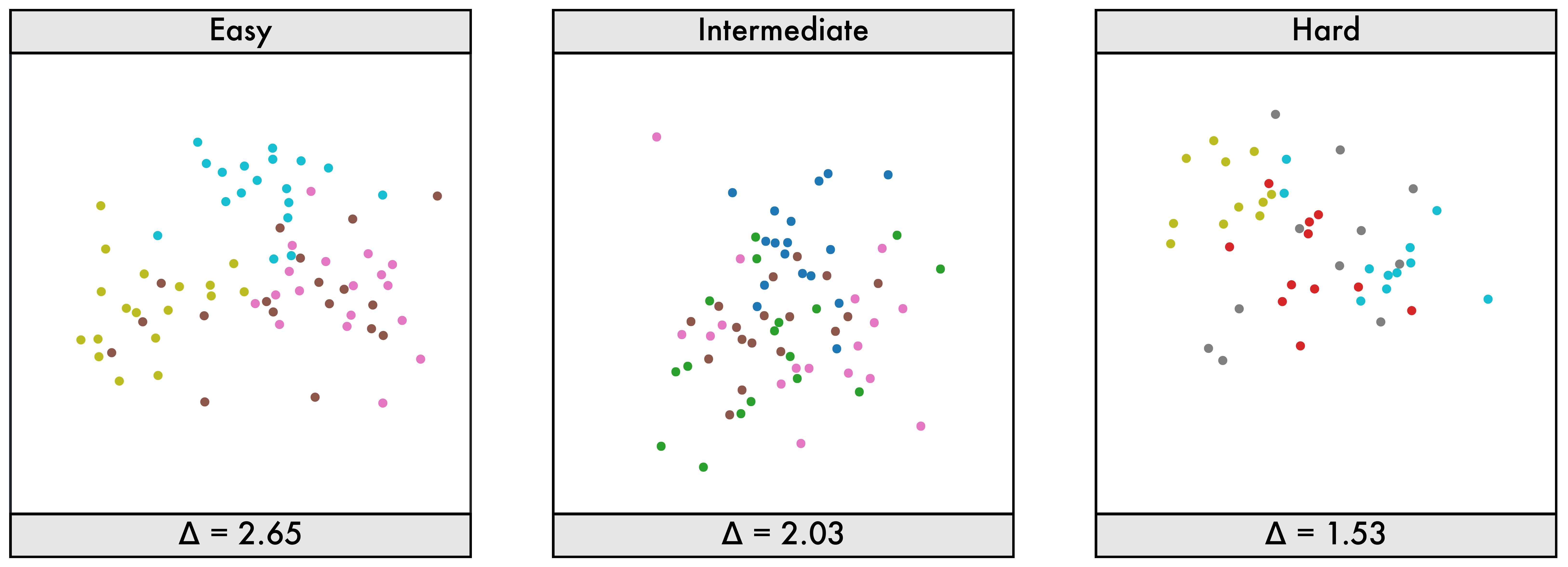} 
\caption{Instances with varying hardness level ($\Delta$) values employed in our study. The difficulty level of instances varies from easy to hard from left to right, with four categories.} 
\label{fig:delta-stimuli}
\end{figure*}


\begin{figure*}[htbp] 
\centering
\includegraphics[width=0.8\textwidth]{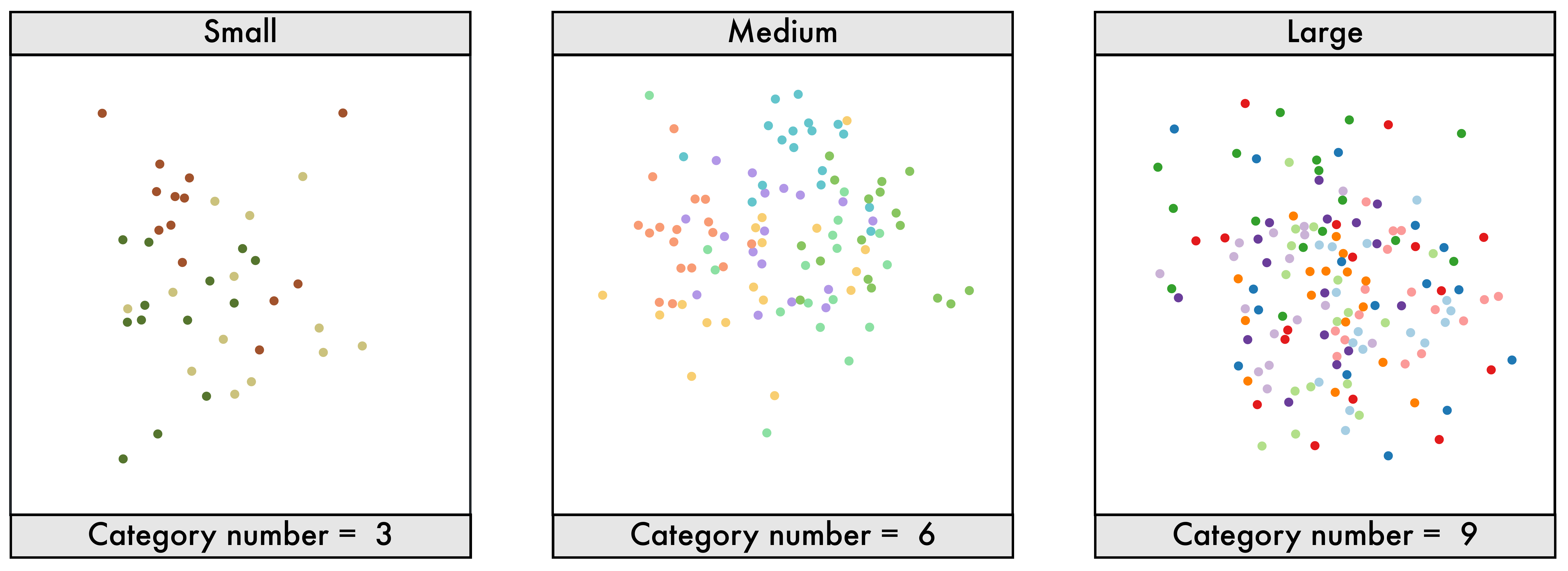}
\caption{Instances with varying numbers of categories employed in our study. Their numbers of categories are 3, 6, and 9 respectively from left to right, with the same hardness level (intermediate).} 
\label{fig:num-stimuli}
\end{figure*}

\subsection{Procedure}
Our experiment consisted of three phases: (1) informed consent and color-blindness screening, (2) task description and tutorial, and (3) formal study.
At the beginning of the study, participants were provided with informed consent in accordance with our IRB protocol. They were then asked to complete an Ishihara test for color-blindness screening~\cite{hardy1945tests}. 
After completing the screening successfully, participants were led to a description page for the mean judgment task. They were required to successfully complete an easy tutorial question to minimize possible ambiguities in their understanding of the task.

During our formal study, each participant completed our target task (\textit{Identify the class with the highest average y-value}) for 45 stimuli presented sequentially using a single color palette (42 formal trials and three engagement checks). We used stratified random sampling to balance number of categories and difficulty levels that each participant saw. 
To ensure participants saw a range of category numbers, we grouped category numbers into three classes: small, medium, and large, which corresponded to 2-4, 5-7, and 8-10 categories, as shown in \autoref{fig:num-stimuli}. 
We also grouped stimuli into three difficulty levels: easy, intermediate, and hard, as shown in \autoref{fig:delta-stimuli}. Each person saw 14 stimuli from each category and difficulty group, with combinations of category and difficulty assigned at random. 

We randomly placed three engagement checks within 42 formal trials to assess if participants were inattentive during the test.
These engagement checks presented three classes with large differences in their means (c.f., Figure~\ref{fig:amt-engagement}). 
We randomly ordered the sequence of the formal questions and the engagement checks to avoid learning or fatigue effects.

\subsection{Participants}
We recruited 95 participants from the US and Canada with at least a 95\% approval rating on Amazon Mechanical Turk (MTurk). We excluded four participants who failed more than one engagement check. We analyzed data from the remaining 91 participants (46 male, 45 female; 24--65 years of age).
All participants reported normal or corrected to normal vision. 
Our experiment took about 15 minutes on average, and each of the participants was compensated \$3.00 for their time.

\subsection{Analysis}


We measured performance as both accuracy and time spent on task.
We analyzed the resulting data using a 10 (color palette) x 9 (number of categories) mixed-factors ANCOVA, with the number of points, interparticipant variation, trial order, and hardness levels as random covariates. 
During our post-hoc analysis, we employed the Tukey's honestly significant difference test (Tukey's HSD) with $\alpha$ = 0.05 and Bonferroni correction. 


\section{Results}
\label{sec-results}

We discuss significant results and statistical analysis based on the independent factors considered in this paper (see Appendix) using both traditional inferential measures and 95\% bootstrapped confidence intervals ($\pm$ 95\% CI) for fair statistical communication~\cite{dragicevic2016fair}. Table \ref{tab:ancova-result} summarizes our ANCOVA results.
Additional results, charts, and details of the analysis can be found on Appendix. 


\begin{figure*}[htbp] 
\centering
\includegraphics[width=0.8\textwidth]{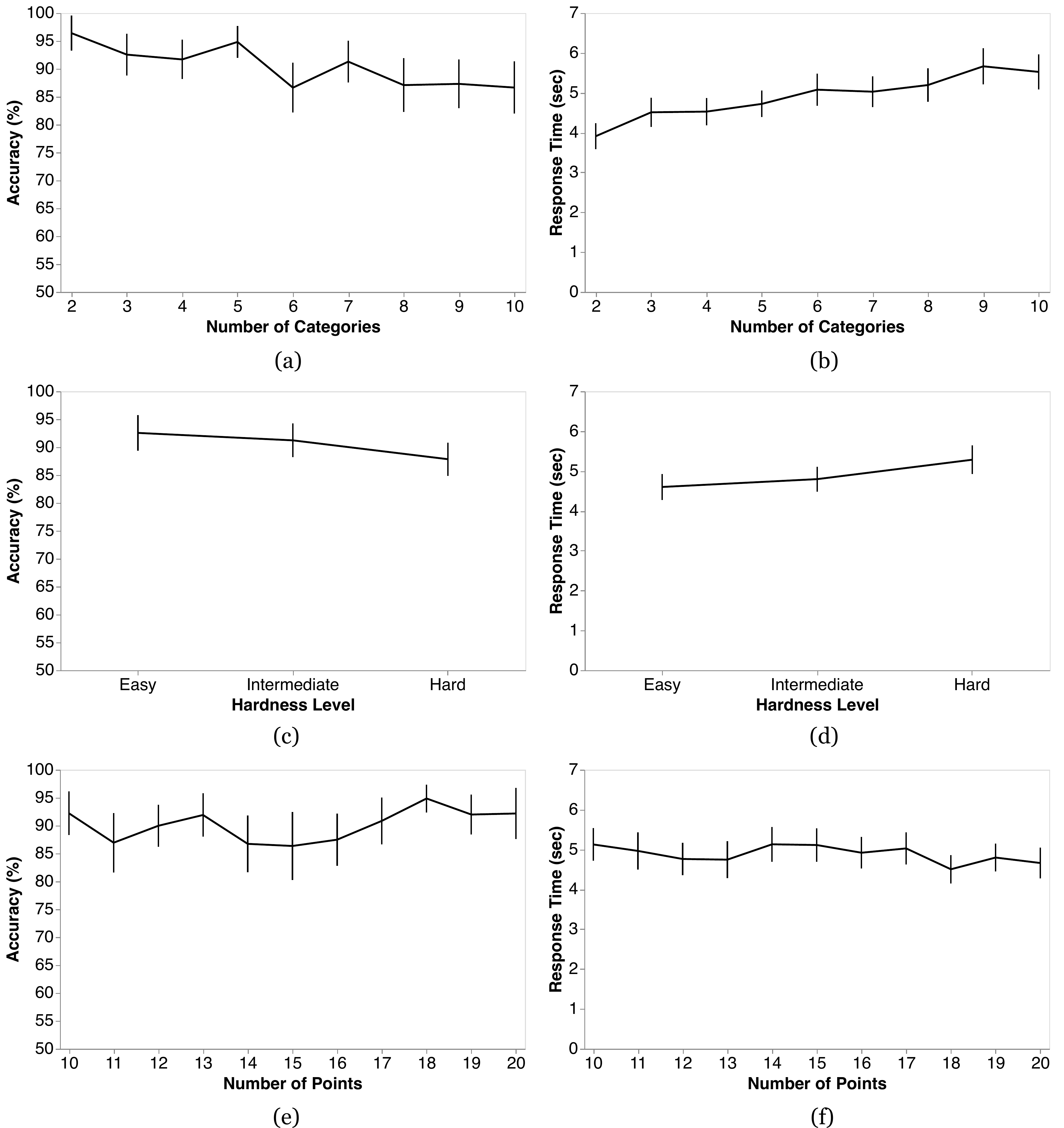}
\caption{Our primary results with respect to the numbers of categories, hardness level, and numbers of points. Graphs on the left show changes in accuracy, whereas those on the right show response times Both accuracy and time do not systematically vary with the number of points. However, as the number of categories grows or the hardness level increases, the overall accuracy rate drops, and the time spent escalates. In order to show the trend clearly, we used a scale from 50--100\% (chance at our smallest number of categories to perfect performance) on the y-axis for accuracy. Error bars represent 95\% confidence intervals.} 
\label{fig:all}
\end{figure*}

\subsection{Number of Categories}
\label{sec-analysis-cat}

Our results support \textbf{H1}:
we found that performance decreased as the number of categories increased. 

Our analysis reveals a significant effect of category number on judgment performance ($F(8, 82)=7.6511, p<.0001$): people were both less accurate and slower with higher numbers of categories.
\autoref{fig:all} (a) shows that accuracy rate decreases based on the number of categories from 96.4\% to 86.6\%, 
with an overall descending trend as the number of categories increases.
\autoref{fig:all} (b) presents the average spent time broken down by category number,
suggesting that participants were slower for scatterplots with more categories.

We also found anomalies in the accuracy rate
for between five and six categories (\autoref{fig:all} (a)).
While we initially assumed this anomaly to be noise, the pattern was repeated across almost all palettes. 
This category number correlates with past findings of \textit{subitizing}---the ability to instantly recognize how many objects are present without counting---in categorical data from Haroz \& Whitney~\cite{haroz2012capacity}. 
While we do not confirm this hypothesis in this study, our results do raise questions about the role of subitizing or a related mechanism in categorical reasoning with visualizations.  

\begin{table}[htbp] 
\centering
\caption{ANCOVA results. Significant effects are indicated by \textbf{bold} text and the corresponding rows are highlighted in green.}
\includegraphics[width=\linewidth]{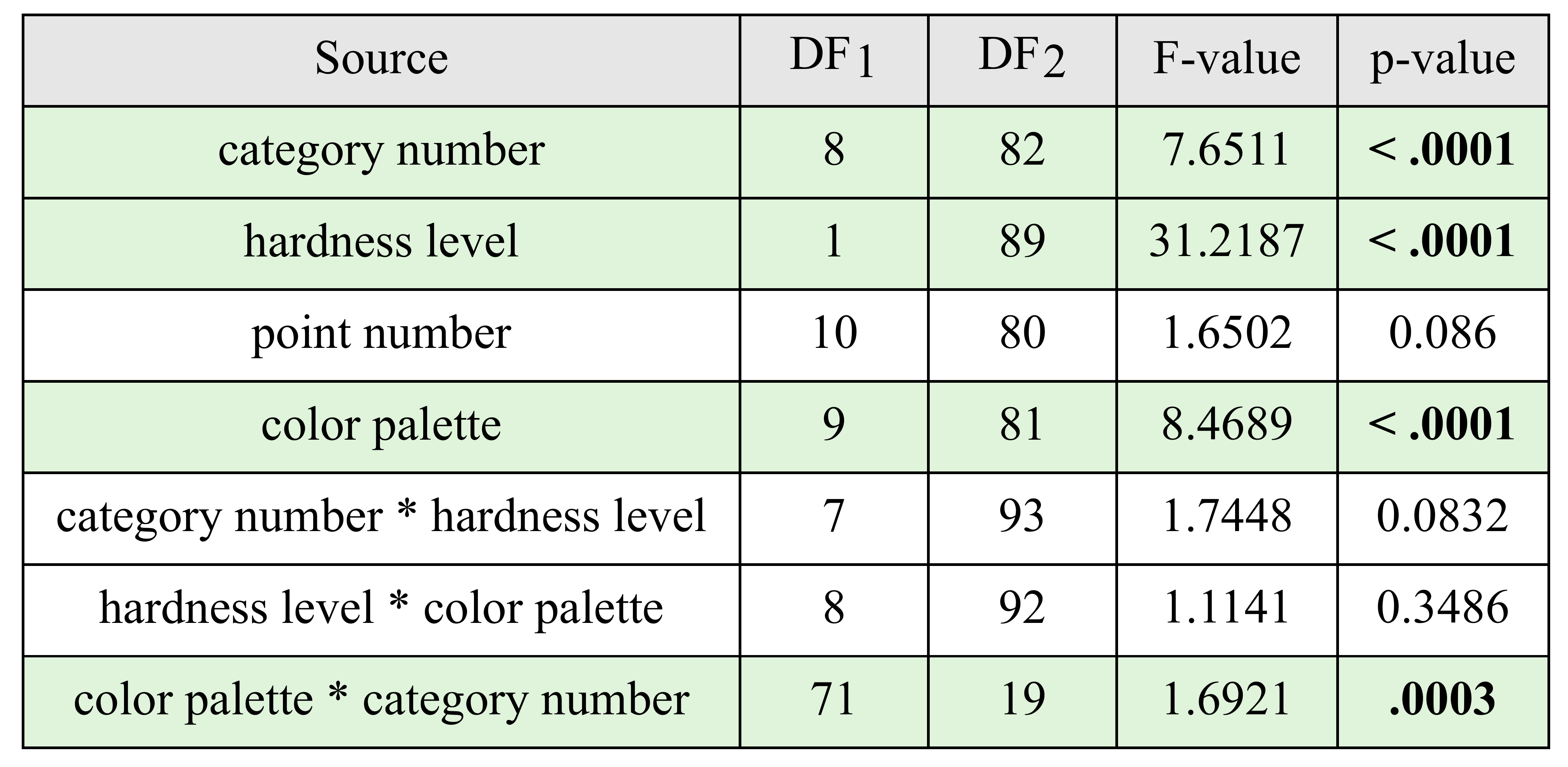} 
\label{tab:ancova-result}
\end{table}

\begin{figure*}[htbp] 
\vspace{-1em}
\centering
\includegraphics[width=0.8\textwidth]{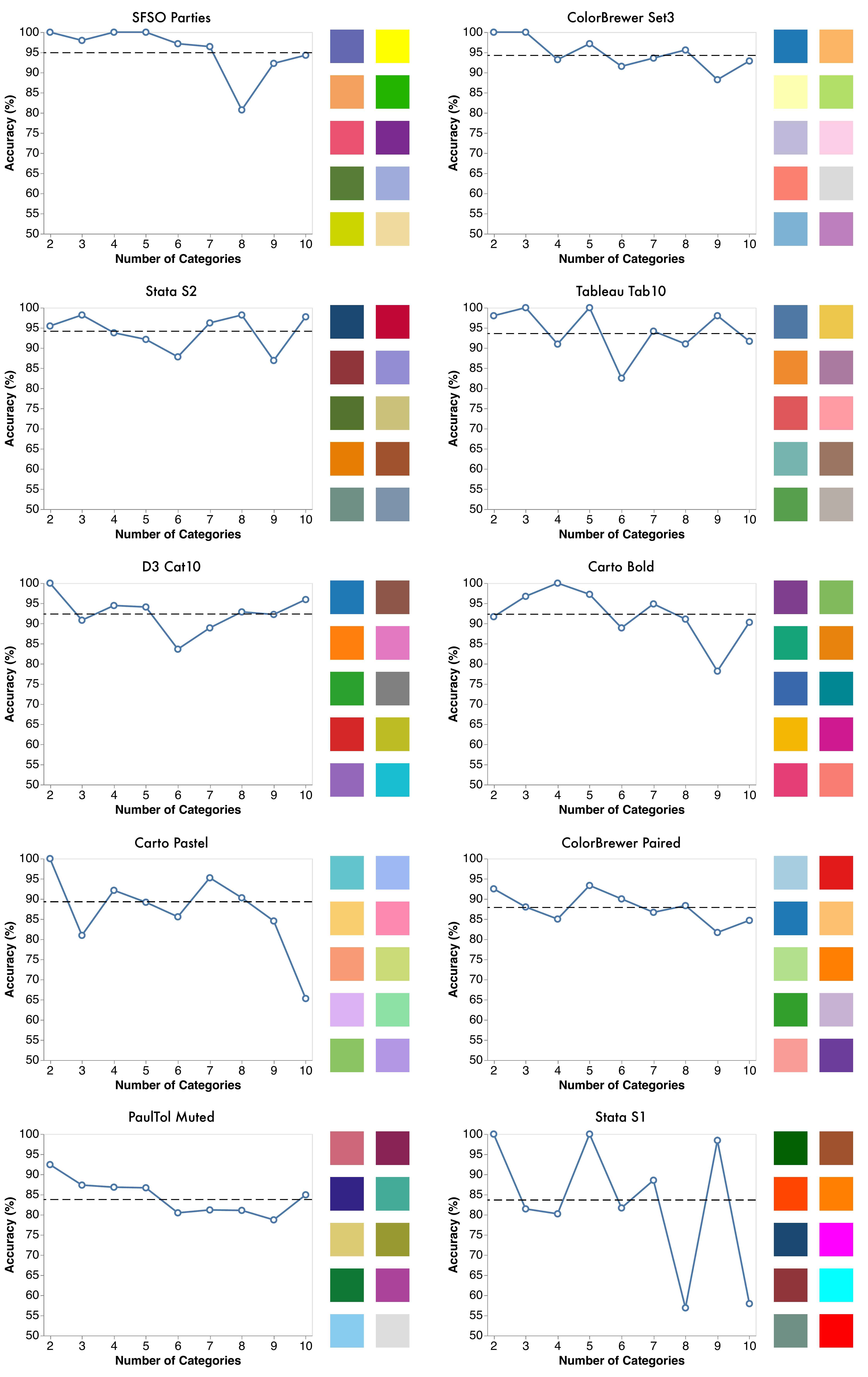} 
\vspace{-1em}
\caption[]{The accuracy rates based on the number of categories separated per color palette, sorted by average accuracy over all categories (dash lines) sorted from most to least accurate. Color palettes are shown along with corresponding charts. See Section \ref{sec-analysis-color} for detailed analysis \add{and \autoref{tab:parameters} in the Appendix for the count of scatterplots per palette.}
} 
\label{fig:palettes-acc}
\end{figure*}

\subsection{Color Palettes}
\label{sec-analysis-color}


Our results also support \textbf{H2}: color palettes significantly affect accuracy ($F(9, 81)=8.4689, p<.0001$, see \autoref{tab:ancova-result}). 
We 
found a significant interaction effect between color palettes and the number of categories for both time and accuracy. 
In other words, as the number of categories increases, the accuracy ranks between color palettes might be different. Different palettes are more or less robust to increasing the number of categories. This finding indicates that there is no best palette
for multiclass scatterplots. Instead, our results provide guidance for designers to select effective palettes based on the number of categories in their data. 

\autoref{fig:palettes-acc} shows the
accuracy rate and category number per color palette. 
These charts reveal that:

\begin{enumerate}
    \item \emph{SFSO Parties} and \emph{ColorBrewer Set3} achieved the highest average accuracy rate in all data, whereas \emph{PaulTol Muted} and \emph{Stata S1} exhibited the worst overall performance (an 11.3\% accuracy difference on average between \emph{SFSO Parties} and \emph{Stata S1}),

\item lower performing palettes tend to be less robust to increasing the number of categories, and 

\item most palettes show an overall descending trend as the number of categories increases, though some palettes remained relatively robust (e.g., \emph{Stata S2}, \emph{D3 Cat10}).

\end{enumerate}




\begin{table}[htbp] 
\centering
\caption{Three performance classes of color palettes from Tukey's HSD. Performance is rated better to worst from Class A to C respectively. }
\includegraphics[width=\linewidth]{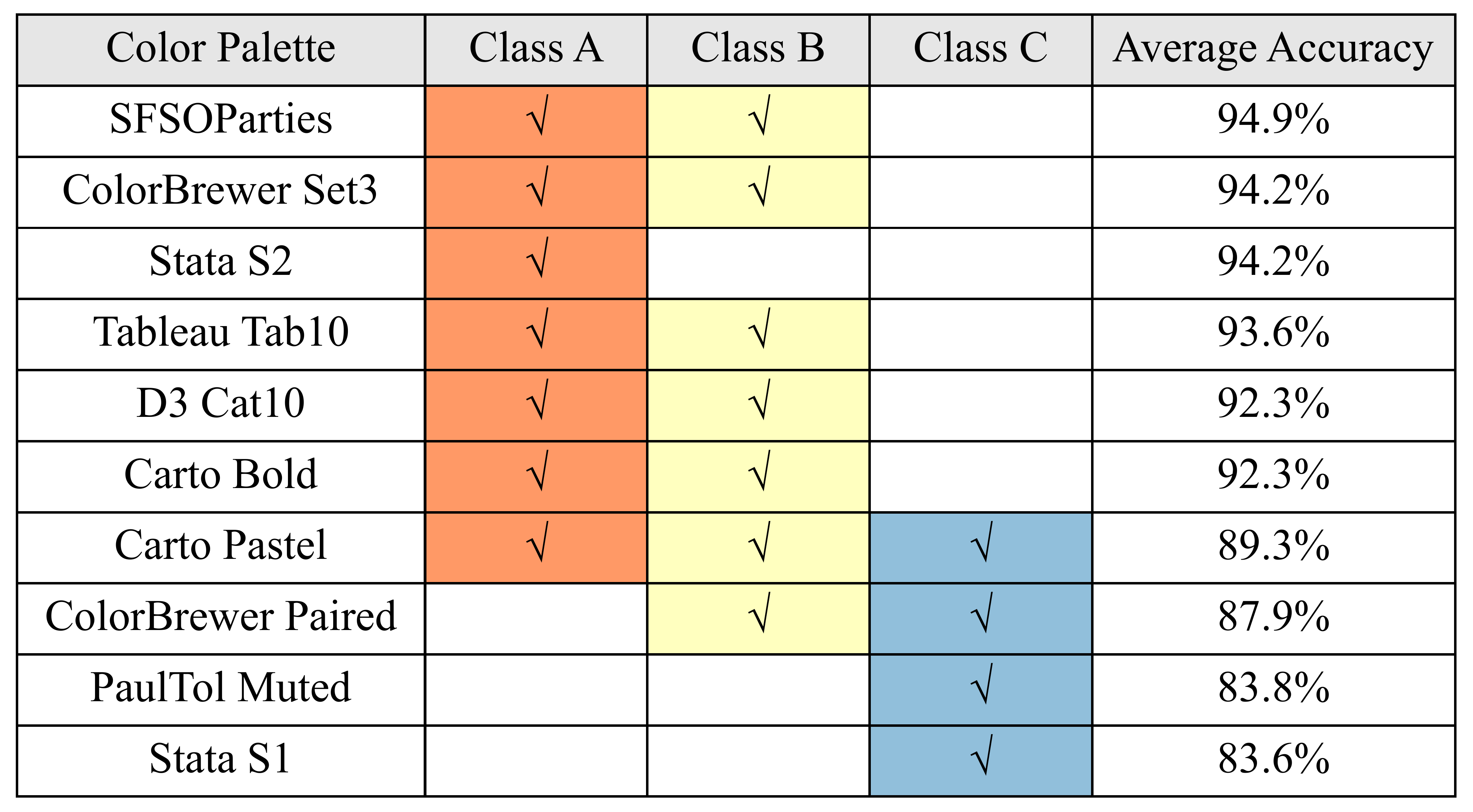} 
\label{tab:color-classes}
\end{table}

\subsection{Exploratory Analysis}
\label{sec-analysis-explor}

\begin{figure*}[htbp] 
\centering
\includegraphics[width=0.95\textwidth]{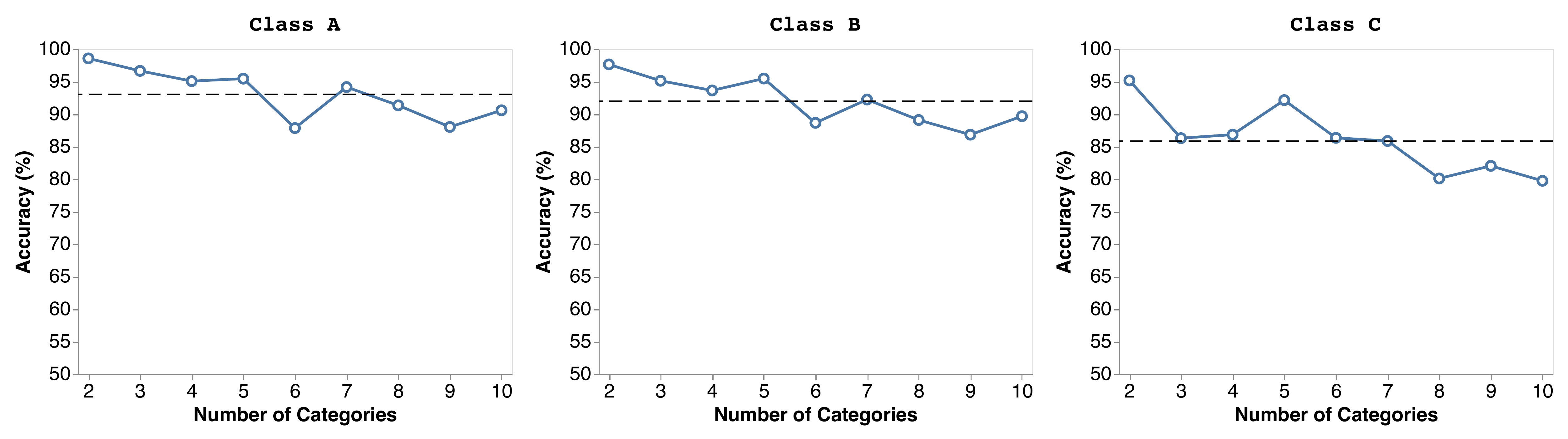} 
\caption{The average accuracy rate with different numbers of categories per performance class of color palettes. Charts represent Class A to C from left to right.} 
\label{fig:class-acc}
\end{figure*}

To better analyze the impact of specific color palettes, we performed a Tukey's HSD with Bonferroni correction to identify significant performance differences between palettes. 
The test revealed three \emph{classes} of color palettes with comparable performance, shown in \autoref{tab:color-classes}.
\autoref{fig:class-acc} illustrates the combined accuracy rate of the three classes, in which Class A refers to the best performance, Class B is slightly lower, and Class C is the worst overall. 
All three classes of palettes showed a steady downward trend that is consistent with \textbf{H1}. We use the clusters created by these performance classes to scaffold an exploratory analysis of potential metrics associated with the observed performance differences. 

We analyzed these classes using eight color metrics associated with palette design to explore the relationship between performance and common design parameters: perceptual distance~\cite{sharma2005ciede2000}, name difference~\cite{heer2012color}, and name uniqueness~\cite{heer2012color} as employed by Colorgorical~\cite{gramazio2016colorgorical} and the magnitude and variances of different dimensions in CIELCh~\cite{zeileis2009escaping} ($L^*$, $C^*$, and $h^*$). 
The 
computations for those metrics can be found in our supplemental materials.
Since we randomly sampled colors in a palette for plots with less than 10 categories (see Section \ref{sec-stimuli-generation}), for each target color palette, we compute those metrics based on the actual colors used in each individual stimulus sampled from the target palette to explore the distribution of these features with respect to performance.

We conducted an ANOVA using these nine measures to assess the impact of each parameter on accuracy (\autoref{tab:metric-significance}).
We found significant effects ($p<0.01$) of $L^*$ variance, $L^*$ magnitude, and all-pairs perceptual distance~\cite{sharma2005ciede2000} and marginal effects ($p<0.10$) of $h*$ variance and $C*$ variance.

\begin{table}[htbp] 
\centering
\caption{Results of significance analysis from color metrics to judgment accuracy.
The right-most column shows plus or minus of the $\beta$-ratio in the OLS linear regression where plus means incremental trend, and minus means decremental trend.
Significant impacts ($p<0.01$) are in \textbf{bold} style and green color. }
\includegraphics[width=\linewidth]{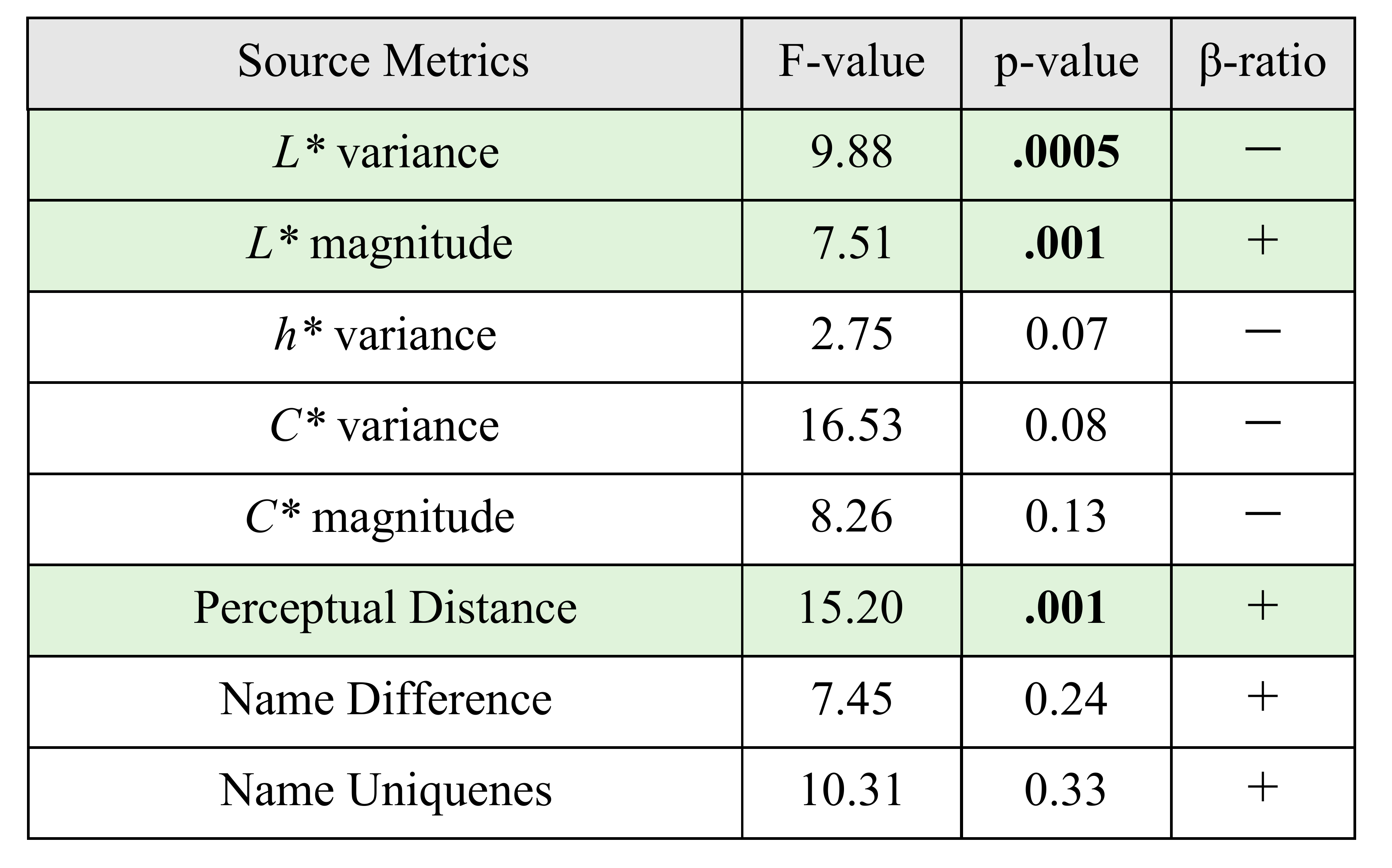} 
\label{tab:metric-significance}
\end{table}

To assess the direction of the effects, we performed an OLS linear regression of each metric and average accuracy. Since the value of $\beta$-ratio (in $Y=\beta X + \epsilon$) of regression differs from the data range of source metrics, we show its 
directionality in the right-most column in \autoref{tab:metric-significance}, where a plus sign refers to an increasing trend and minus sign means decreasing trend.
We found that larger $L^*$ magnitude (lighter colors) and larger perceptual distance both lead to better judgment accuracy, whereas palettes with less lightness variance had better performance. 

Our findings suggest that palettes leveraging lighter colors, larger perceptual distances, and lower luminance differences led to better performance \add{whereas factors like hue variance or name uniqueness that are conventionally associated with palette design may be less important when considered in isolation}.
The use of lightness in palettes has a tenuous history: some advocate for minimizing lightness variation to avoid biasing attention \cite{munzner2014visualization,kindlmann2002face}, whereas other  guidelines suggest that designers leverage higher contrasts introduced by lightness variations to take advantage of our sensitivity to lightness variations~\cite{rogowitz2001blair}.
Our results provide further support for privileging more isoluminant palettes to avoid directing too much attention to given classes
\cite{braithwaite2010visual}.
However, future work should more systematically explore this hypothesis.

\section{Discussion}
\label{sec-discussion}
We measured the impact of the number of categories and choice of color palettes on people's perceptions of multiclass scatterplots.
We find that people are less accurate at assessing category means as the number of categories increases. However, certain palettes may be more robust to variations in category number. 
Our results provide new perspectives on prior findings and offer both actionable design guidance and opportunities for future research. 

\subsection{Reflections on Prior Findings}






Our study indicates that increasing numbers of categories would lead to descending performance in relative mean estimation (see Section \ref{sec-analysis-cat}).
However, this finding is inconsistent with the proposed guidelines of Gleicher et al.~\cite{gleicher2013perception}, which found little impact of adding additional distractor classes. We anticipate that this contradiction is a result of the number of classes evaluated. For smaller numbers of classes, performance tended to be more robust across palettes. However, as the number of classes increased, performance started to degrade. This finding likely stems from a correlation between performance and color discriminability indicated in our exploratory analysis. As the number of colors increases, it becomes more difficult to ensure colors are spaced apart, especially if maximizing for metrics like pair preference~\cite{schloss2011aesthetic} or otherwise minimizing the number of large hue variations to preserve harmonies \cite{stone2006choosing}. Certain palettes may differently balance this aesthetic and performance trade-off. However, our results indicate that this trade-off is sensitive to the number of categories present in the data. 

\add{Contrary to past heuristics, we found that performance remained relatively high even for more than seven categories. We hypothesize that people may simply be better at this task than they expect: while the task may feel significantly more difficult as the number of categories increases, our visual system may be more robust than expected in working with complex categorical data. Feedback from pilot participants indicated that the tasks felt challenging as the number of categories exceeded four, but these participants, like those in the formal study, still performed well at such seemingly difficult tasks. 
While the tested palettes reflect best practices, our findings challenge existing heuristics around the scalability and utility of color in visualization. A more thorough and rigorous empirical examination of the robustness of categorical perception in visualization generally would benefit a wide range of applications.}

\add{Our results demonstrate a high overall accuracy (more than 85\%) compared to Gleicher et al. ~\cite{gleicher2013perception} ($\sim$ 75\%) even though we tested a larger number of categories. 
Gleicher et al. chose to generate uniformly sparsely distributed classes. 
However, such distributions might not be widely applicable; in real-world use cases, 
people more commonly build insight from densely distributed classes~\cite{van2008visualizing}.
As shown in \autoref{fig:num-stimuli}, we generated more randomly and densely distributed classes to privilege ecological validity.
Consequently, our stimuli are more likely to perform similarly to what people usually see in their daily life, but the clustering structure may have affected task accuracy by, for example, making it slightly easier to group points within categories.
While we tuned our stimulus difficulty in piloting and our results were consistent across different performance thresholds, providing evidence of their generalizability, these differences raise important future questions as to the impact of different data distributions on categorical palette design. }



Part of the difference in results between our findings and Gleicher et al. may stem from differences in perceptual mechanisms present when processing different numbers of categories. 
Our study reveals a significant "dip" in accuracy when the number of categories increased to five or six (see Section \ref{sec-analysis-cat}).
These bumps correlate with a key number of objects for subitizing~\cite{kaufman1949discrimination}: below roughly six objects, we can instantly and precisely detect the quantity of objects present, whereas we have to actively count larger numbers. While subitizing tends to focus on individual objects rather than collections of objects, the dip in accuracy directly correlates with this subitizing threshold and echoes similar findings in past work in categorical visualization~\cite{haroz2012capacity}. Our study is not designed to probe subitizing or other specific perceptual mechanisms that may explain these results. However, this correlation offers opportunities for further understanding the relationship between categorical perception, subitizing, cluster detection, and other related perceptual phenomena in visualization. 

We also found a significant overall difference between color palettes. These differences echo the findings of Liu \& Heer \cite{liu2018somewhere} for continuous colormaps: even if a palette satisfies the basic constraints of good palette design (e.g., discriminable colors), it may not perform optimally. Like Liu \& Heer, we also find that characterizing the source of these performance differences is challenging: palette effectiveness arises from a complex combination of factors. Future work should seek to further deconstruct these factors to derive more robust design guidelines. 

\subsection{Design Guidelines for Multiclass Scatterplots}

The data and design 
of multiclass scatterplots significantly influence our abilities to reason across classes. 
Compared to Gleicher et al.'s guidelines~\cite{gleicher2013perception}, our results emphasize the 
influences of category number and 
color palette, which are the two essential elements in visualizing categorical data. 
Additionally, in contrast of some existing guidelines for color palettes~\cite{ware1988color, rogowitz2001blair}, our results indicate maximizing luminance variation may hinder analysis. While designers can use our results to directly choose the optimal palette from our tested set of palettes given the number of categories in their data, our results also provide preliminary guidance for palette selection more broadly: 

\vspace{3pt}\noindent \textbf{Simplifying category structure may improve performance.}
Our study suggests that people can reason across multiple classes encoded using color.
However, as shown in Section \ref{sec-analysis-cat}, designers should be aware that performance tends to degrade as the number of categories increases: people are slower and less accurate, especially when working with six or more categories.
We recommend designers 
consider how the number of categories influences performance on key tasks and consider collapsing relevant categories hierarchically if necessary. 

\add{As a caveat, people were relatively good at completing this task, even with larger numbers of categories than conventional heuristics recommend. Our results indicate that people can reliably distinguish colors in large palettes even though informal pilot participants indicated that the task felt quite difficult for higher numbers of categories. This contrast between perceived and objective performance suggests that even well-established design heuristics can benefit from experimental validation and refinement.  }

\vspace{3pt}\noindent \textbf{When designing new palettes,  consider fewer lightness differences, larger perceptual distances, and lighter colors.}
As shown in Section \ref{sec-analysis-color}, our results reveal that color palettes significantly impact the accuracy of human judgment.
Our exploratory analysis confirms the benefits of maximizing the pairwise difference between colors and provides further evidence of minimizing lightness variation. However, we also find that palettes using lighter colors tend to also enhance accuracy. We anticipate that this bias may be in part due to the use of a white background enhancing contrast within categories while minimizing undesirable ``loud'' colors that have too high of a luminance contrast with the background. 
However, the tested palettes are all handcrafted to select harmonious and aesthetically pleasing colors. Future work should investigate 
these results 
on other background colors. \add{We also found little evidence of the benefits of hue or color name variation when considered in isolation. This points to the need for the systematic interrogation of designer practices to improve existing heuristics for palette design \cite{smart2019color}.}


\vspace{3pt}\noindent \textbf{Choose your palettes to fit your data.}
When the number of categories changes, the performance rank of different color palettes may also change. Different palettes are differently robust to changes in category number. We recommend designers select color palettes based on the parameters of their specific data.
For example, a designer might use \emph{SFSO Parties} or \emph{ColorBrewer Set3} for multiclass scatterplot with less than seven categories and \emph{D3 Cat10} for larger numbers of categories (see \autoref{fig:palettes-acc}).

\subsection{Limitations and Future Work}
We studied the impact of the number of categories and color palettes on multiclass scatterplots. 
However, scatterplots offer a wide variety of design choices for representing categorical data that may
provide different trade-offs in perception~\cite{sarikaya2018scatterplots}. 
Future work should explore the robustness of different channels to varying numbers of categories. Further, scatterplots often encode larger numbers of variables, such as multiple categorical dimensions or combining categorical and continuous dimensions~\cite{smart2019measuring}. Future work should investigate the interplay between different design factors in higher dimensional multiclass scatterplots. 
\add{Both our study and Gleicher et.al.'s work~\cite{gleicher2013perception} focused on 
comparing y values. However, scatterplots are two-dimensional visualizations. Future work should consider the impact of palettes on crossdimensional tasks.}

We 
evaluated 10 pre-defined qualitative color palettes on qualitative data.
\add{We employed a random color sampling strategy from selected palettes for 
data with less than ten categories to simplify the stimulus generation to avoid potential bias from sources outside of color selection. Future work should extend our results to consider sequential strategies in comparing preconstructed palettes.}
Additionally, categorical data can also be encoded using other types of palettes, such as sequential and diverging encodings~\cite{liu2018somewhere}, whose robustness to varying numbers of samples is not well understood. 
\add{
Considering additional properties of color selection, such as accessible palettes for people with color vision deficiencies~\cite{jefferson2006accommodating, pugliesi2011cartographic}, is also important future work.
}

We sampled from predefined palettes at a fixed mark size. Varying mark size can influence mark discriminability~\cite{szafir2018modeling}. As mark size was held constant for all palettes and all palettes had large distances between all color pairs, we do not anticipate that this choice biased our results. However, future  work should explore a larger range of mark sizes and mark types. It should also seek to more systematically evaluate the robustness of our exploratory results. Such variation is challenging due to a large number of potential perceptual factors; however, our results may provide preliminary support for identifying the most promising factors. 

\add{We elected to use Mechanical Turk to reflect the range of viewing conditions and participants common to web-based visualizations and to recruit larger numbers of participants. However, variations in viewing conditions can influence color perception. While past studies of color perception in visualization validate the predictive ability of crowdsourced studies for color perception studies in HCI \cite{szafir2014adapting,reinecke2016enabling}, the variability introduced by the range of viewing conditions on MTurk limits the generalizability of our results and our ability to make precise claims about fine-grained mechanistic perceptual phenomena. However, given the large differences between colors in our palettes, we anticipate the affect of viewing variation to be relatively minimal \cite{moroney2003unconstrained} and followed best practices in our experimental design to minimize the impact of viewing variation. Future work seeking to quantify more precise causal mechanisms underlying our findings may wish to replicate our study under more constrained conditions. 
}

Additionally, data-centric statistical factors that may be related to the performance of multiclass scatterplots are not considered in our study.
For example, we did not explore the impact of correlation or strength of clusters.
Extending our experiments to consider a wider range of data properties as well as statistical tasks 
could help us further understand categorical data visualization for complex datasets and usage scenarios and offer broader guidance for categorical visualization generally. 

\section{Conclusion}
\label{sec-conclusion}

We measure how different color palettes impact people's ability to distinguish classes and assess mean values on multiclass scatterplots.
Our results suggest that both the number of categories and the discriminability of color palettes heavily impact people's abilities to use multiclass scatterplots.
We found that increasing the number of categories decreases how well people can distinguish different classes. Furthermore, we found preliminary evidence that even using designer-crafter palettes, a more discriminable color palette (such as \emph{SFSO Parties} who achieves 95\% average accuracy) can perform nearly 12\% better than a less discriminable one (such as \emph{Stata S1} with only 83\% average accuracy).
Based on the experimental results, we critically reflect on past findings and derive a set of design guidelines for palette selection in multiclass scatterplots.
We believe that our findings have the potential to support a variety of other visualization types and low-level tasks that combine continuous and categorical data. We hope our work will inform future studies to construct more general guidelines for the understanding of categorical perception in information visualization.

\begin{acks}
This work was supported by NSF IIS \#2046725 and by NSF CNS \#2127309
to the Computing Research Association for the CIFellows Project.
\end{acks}

\bibliographystyle{ACM-Reference-Format}
\bibliography{main}

\appendix
\section{Appendix: Pilot Study}
\label{sec:appendix}

Our study aims to understand the robustness of color palettes on the perception of multiclass scatterplots.
To tune the parameters of our study, we first conducted three pilot studies to identify 
people's abilities to recognize data about different visual factors in multiclass scatterplots and to decide the proper parameters for scatterplots in stimuli generation.

\subsection{Factors}
\label{sec-factors}
We first describe the independent visual factors 
we considered for generating multiclass scatterplots in both pilot and formal studies.

\textbf{Number of categories.}
The total category count in a scatterplot, varies from 2-10 in our experiments.

\textbf{Level of difficulty.}
We described the distance of means between the categories that have the highest mean and the second highest mean to be $\Delta$. We considered the task to be easier as the $\Delta$ is larger, and more difficult as the $\Delta$ is smaller. 

\textbf{Point distribution.}
The pre-generated x-y data of points. Points from each category were randomly sampled from the Gaussian distribution.

\textbf{Number of points.}
The number of points in one category. Each category in the same scatterplot shared the same number of points, varying from 10-20. 

\textbf{Color palettes.}
10 color palettes in total were used in our experiments, with 10 colors in each palette. A certain number of colors were randomly picked to display in each scatterplot depending on the number of categories.

\subsection{Procedure}
We followed the same procedure in the three pilot studies.
Participants were required to carefully read the task description first and then 
completed a tutorial check to ensure their understanding.
Afterward, for each study, all participants viewed scatterplots from the corresponding dataset to make it a fair comparison.
They were required to pick the class with the highest average y-value.

\begin{table*}[htbp] 
\centering
\caption{\add{The number of samples collected for each experimental condition after exclusions. Columns are category numbers and rows are color palettes.  }} 
\includegraphics[width=0.9\textwidth]{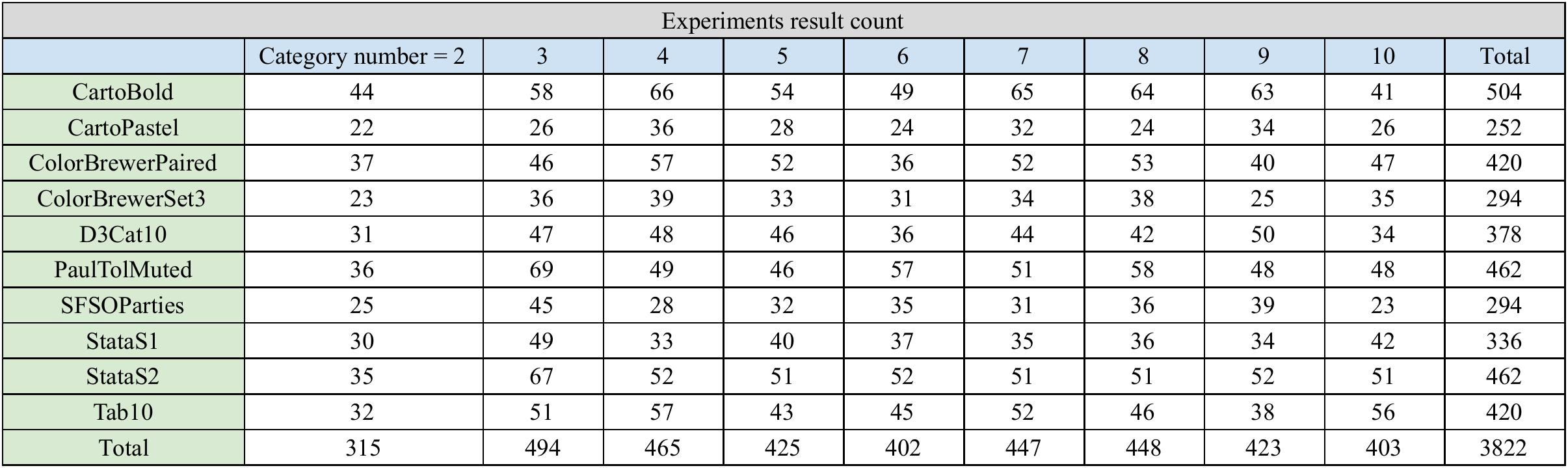} 
\label{tab:parameters}
\end{table*}

\subsection{Pilot Study 1: Hardness Level ($\Delta$) of Stimulus}

\textbf{Participants.} We recruited 106 participants for this study. Participants are all college students, other demographic information was not recorded. They all participated voluntarily and no compensation was provided. 

\textbf{Generation factors:}
Number of categories: \{2\};
Level of difficulty: \begin{math} \{ \Delta \in \mathbb{R} \, | \, 0.5<\Delta<5 \} \end{math};
Point distribution: Poisson distribution with data points (x,y) denoted as \begin{math} \{ x, y \in \mathbb{R} \, | \, 0<x,y<10 \} \end{math};
Number of points: \{15\};
Color palettes: \emph{D3 Cat10}.

\textbf{Results.} The overall accuracy of this study is 76.88\%.
The results suggested that the accuracy rate will increase with the $\Delta$ rises.
To avoid showing tasks that are too easy or too difficult for participants, 
we selected $\Delta$ from 1.5 to 3.0 in the final study. In the formal study, we mark the $\Delta$ in range 1.5 - 2.0 as hard level, 2.0 - 2.5 as intermediate, and 2.5 - 3.0 as easy (c.f., \autoref{fig:delta-stimuli}).
Details of the result and figures are available in our supplemental material.

\subsection{Pilot Study 2: Number of Categories} 

\textbf{Participants.} We conducted the second study with 25 participants from the 
UNC campus. Other demographic information was not recorded. They all participated voluntarily and no compensation was provided. 

\textbf{Generation factors:}
Number of categories: [2, 9];
Level of difficulty: \begin{math} \{ \Delta \in \mathbb{R} \, | \, 1.5<\Delta<3.0 \} \end{math}
Point distribution: Poisson distribution with data points (x,y) denoted as \begin{math} \{ x, y \in \mathbb{R} \, | \, 0<x,y<10 \} \end{math};
Number of points: \{5, 10, 15\};
Color palettes: \emph{D3 Cat10}.

\textbf{Results.} The overall accuracy of this study was 98.30\%.
The result revealed that participants can identify mean judgment across a lot of categories and colors.
Likewise, we decided to use 2 to 10 categories in the final study, 
see \autoref{fig:num-stimuli} for examples.
The extremely high accuracy rate encouraged us to think about whether the results are impacted by our choice of distribution. We conducted a third study to check if the Poisson distribution is too na\"ive for this task.
Details of the result and figures are available in our supplemental material.

\subsection{Pilot Study 3: Point Distribution}

\textbf{Participants.} 81 participants joined the third study in total. All the participants were recruited from Amazon Mechanical Turk (MTurk), aged between 24 to 65, with an average of 37 with a standard deviation of 10.7. There are 51 males and 30 females, and 69 of them are wearing corrected glasses.

\textbf{Generation factors:}
Number of categories: [2, 10];
Level of difficulty: \begin{math} \{ \Delta \in \mathbb{R} \, | \, 1.5<\Delta<3.0 \} \end{math}
Point distribution: Gaussian distribution with data points (x,y) denoted as \begin{math} \{ x, y \in \mathbb{R} \, | \, 0<x,y<10 \} \end{math};
Number of points: [10, 20];
Color palettes: All 10 color palettes, see \autoref{fig:palettes-acc}.

\textbf{Results.} The overall accuracy of this study was 80.10\%.
The result suggested that there might be a cue between category number and human judgment accuracy.
Compared to the Poisson distribution in Pilot Study 2, the accuracy rate 
did not suggest a risk of ceiling effects.
As a result, we decided to use Gaussian distribution to generate scatterplots in our final study.
Details of the result and figures are available in our supplemental material.

\subsection{\add{Metadata}}

\add{\autoref{tab:parameters} illustrates 
the distribution of collected data samples} counted by color palettes and category numbers. Conditions were assigned based on stratified random sampling as described in Section \ref{sec-methodology}.

\end{document}